# Physics of Electrolytic Gas Evolution

C. A. C. Sequeira*, D. M. F. Santos, B. Šljukić, and L. Amaral

*Department of Chemical Engineering, Instituto Superior Técnico, Technical University of Lisbon (TU Lisbon), Av. Rovisco Pais, 1049-001 Lisboa, Portugal*

\* E-mail: cesarsequeira@ist.utl.pt (C. A. C. Sequeira); Tel: +351 218417765

A brief analysis of the physics and effects of electrolytic gas evolution is presented. Aspects considered include bubble nucleation, growth, and detachment, enhancement of mass and heat transfer, and decrease of apparent electrical conductivity of bubble containing electrolytes. This analysis is mainly oriented to hydrogen/oxygen evolving electrodes.

*Keywords: bubble nucleation, growth and detachment; hydrogen-evolving electrodes; oxygen-evolving electrodes; alkaline electrolysis; electrolyte conductivity; mass transfer.*

## 1 INTRODUCTION

Electrolytic gas evolution process is of great importance in numerous electrochemical reactions and devices. These include primary electrode reactions of alkaline water and chlorine electrolysis, side reactions during charging of lead-acid batteries, metal electrowinning, anode reaction of Hall process for aluminium production, chlorate production etc. In fact, one could say that there are relatively few main electrochemical processes in which evolved gases do not appear.

At present, we are concerned with the electrolytic hydrogen ($H_2$) production, which involves two important gas evolution processes, the hydrogen evolving reaction and the oxygen ($O_2$) evolving reaction. Clearly, the kinetics of these reactions is very dependent on the physics and effects of the evolved gases, which justify the present analysis.

Electrolytic gas evolution study comprises two subjects: its physics (the phenomenon by which bubbles grow) and its effects (the havoc engendered by their presence). The physics of gas evolution proceeds through three phases: nucleation, growth and detachment. Bubbles start nucleating at electrode surface from solution once the solution becomes highly supersaturated with produced gas. Subsequently, bubbles grow by dissolved gas diffusion to their surface or by coalescence with others, and finally detach from the electrode when the forces pulling them away overcome the surface forces binding them. Many phenomena of gas evolution within each of these stages and the effect of process parameters on the stages have been theoretically and experimentally investigated, but much remains still to be done. Just as growth of bubbles is complicated, so are the effects of gas bubbles which include electrolyte solution mixing and obstruction of electric current. Impeding the current, bubbles decrease electrolyte conductivity and hence add to ohmic losses in the cell. Furthermore, mixing the electrolyte, bubbles enhance heat transfer from the electrode and mass transfer to the electrode surface. In this study we will summarise the physics and effects of electrolytic gas evolution, being mainly concerned with the presence of bubbles in alkaline water, particularly close to both the hydrogen and oxygen evolving electrodes.

# 2 BUBBLES NUCLEATION

Bubbles may nucleate when the electrolyte near the electrode is supersaturated with gas as, for example, in the electrolytic production of $H_2$. A flux of $H_2$ based on a current of a few $mA/cm^2$ is sufficient to supersaturate the liquid since $H_2$ is sparingly soluble. When the dissolved gas concentration reaches a critical value, bubbles nucleate and grow. The critical dissolved $H_2$ concentration leading to nucleation can be theoretically obtained from classical nucleation theory; however, the effectiveness of this theory for electrolytic gas evolution needs further examination.

Nucleation theory dates from the 1920s when investigators considered the thermodynamics and kinetics of single-component phase transitions [1-8]. From this work evolved classical nucleation theory in which density fluctuations engender vapour nuclei that may grow or decay depending on whether the bubble nucleus is higher or smaller than a certain size, determined by so-called critical radius. At this stage, the bubble is in metastable chemical and mechanical equilibrium with its surroundings. The frequency of formation of critical nuclei is inversely proportional to the exponential of the isothermal minimum work required to form such nuclei divided by kT, being k the Boltzmann constant and T the temperature. Thus, the essential features of nucleation theory are the expression defining the size of the critical bubble nucleus and the formulation of the rate expression as an exponential function of the work associated with the production of a nucleus having this dimension. Blander and Katz [9] have reviewed the derivation of these equations.

Electrolytically evolved gas bubbles nucleate from solutions of gas dissolved in a host liquid; hence theory for multiple components is required for this case. Ward [10] obtained an expression for the critical radius for multi-component systems:

$$R_c = \frac{2\sigma}{\dfrac{\eta P\infty}{v_1} + \dfrac{P'C'}{v_2 \underline{C}_0} - P'}$$ (1)

where P∞ is the vapour pressure of solvent at the temperature of the liquid, P′ is the external pressure in the liquid, $v_1, v_2$ are the vapour phase activity coefficients of the solvent and solute, C′ is the concentration of the gas in the solution surrounding the bubble, expressed as moles of solute per mole of solvent, $\underline{C}_0$ is the equilibrium concentration of the gas in the solvent when a flat surface of the solvent is exposed to the gas only at T′ and P′, expressed as moles of solute per mole of solvent, σ is the surface tension of the liquid-gas interface often assumed equal to the surface tension of the liquid vapour interface, and η is defined as

$$\eta = \exp\left(\frac{V_1(P'-P\infty)}{kT} - C'\right)$$ (2)

where $V_1$ is the specific volume of the pure solvent.

Tucker and Ward [11] derived a general expression for the critical radius of a bubble for an arbitrary number of volatile solutes present. Their final expression has the same form as (1), but the partial pressure in the denominator is summed over all solutes. The solute gas decreases the critical radius from that of the pure solvent at the same temperature and pressure; therefore the rate of nucleation correspondingly increases. Ward *et al.* [10] have also shown that the rate equation derived for the pure solvent,

$$J = Z \exp\left(\frac{-16\pi\sigma^3}{3kT(P''-P')^2}\right)$$ (3)

(where J is the frequency of nucleation $(s \cdot cm^3)^{-1}$, Z is the frequency factor, a very weak function of T and P, and $P''$ is the pressure inside the critical size bubble) also applies to multi-component solutions; however, $P''$ has two contributions in this case. The dissolved gas exerts a vapour pressure in addition to the vapour pressure of the pure solvent. The increase of $P''$ lowers the critical radius and therefore decreases the maximum sustainable supersaturation.

Equation (3) and equation (4) are sufficient for the prediction of the maximum attainable limit of supersaturation.

$$P'' = \frac{P'C'}{v_2 \underline{C}_0} + \frac{P\infty}{v_1}\left[\exp\left(\frac{V_1(P'-P\infty)}{kT} - C'\right)\right] \qquad (4)$$

On the right side of equation (3) is the exponential function that governs the rate of nucleation while equation (4) relates the pressure inside the critical size bubble to the concentration of dissolved gas in the surrounding liquid.

Supersaturation of $H_2$ gas dissolved in 1 N sulphuric acid ($H_2SO_4$) evaluated using equations (3) and (4) is found to be a thousandfold as the limit [12]. However, supersaturation in the vicinity of gas evolving electrodes does not reach this magnitude; instead supersaturations of $O(10^2)$ were found in aqueous solutions. The reason is that these theories do not take into account nucleation at solid surfaces that unavoidably have imperfections that work as nucleation centres [13,14]. It was reported that nucleation on microelectrodes occurs at pits and scratches as preferred sites [15] or, in the case of rotating platinum (Pt) wire, at specific sites that depend on pre-treatment as well as on current density [16]. Furthermore, study of $H_2$ evolution in $H_2SO_4$ solutions on mercury (Hg) as an ideally smooth electrode [12] (with Hg electrode area core for nucleation being continuously renewed) showed that nucleation at supersaturations was much less than predicted by theory. This indicated a possible effect of the strong electric field within the electric double layer. Later studies from theoretical models and experimental results described in the literature have cast new light on the underlying mechanisms of the nucleation of electrolytic gas bubbles [17-22].

## 3 GROWTH STAGE

The growth stage of gas evolution includes the diffusion of dissolved gas to the gas/liquid interface and the coalescence of bubbles. Scriven [23] and others [24,25] theoretically analysed the mass transfer of dissolved gas to the gas/liquid interface. Scriven's [23] square-root-of-time growth dependence was experimentally confirmed by Westerheide and Westwater [15] for a single bubble, while multiple bubbles interfered with each other. Still, this study established diffusion of gas to the bubble surface as at least the initial mechanism by which bubbles grow. Glas and Westwater [26] extended the study including different gases and different electrode materials.

The diffusion-limited initial stage of growth being established, significance of coalescence in gas evolution was pointed out. Janssen and von Stralen [27], exploring $O_2$ bubbles evolution in aqueous potassium hydroxide (KOH) on a transparent nickel (Ni) electrode, found frequent coalescence, mobility of bubbles on the electrode surface, a radial movement of small bubbles toward large ones and consumption of small bubbles by large ones at high current density. Based on observation of $H_2$ bubbles production in acid media, Putt [28] suggested that bubbles grew large by a scavenging mechanism, i.e. by sliding along the electrode surface and consuming other smaller bubbles. On the basis of high speed motion pictures taken from the backside of a transparent electrode, Sides [29] proposed a cyclic mechanism of bubble growth in the basic medium. This mechanism starts with nucleation, followed by growth by diffusion, coalescence of small bubbles, radial movement of small bubbles and their coalescence with stationary medium-size bubbles, and scavenging coalescence of the medium bubbles by large ones moving along the electrode. This is the process by which large $O_2$ bubbles are built in basic medium.

A film sequence recording this mode of coalescence appears in Fig. 1 [30].

FIG 1

The cyclic process of bubble growth has been seen on a much larger scale by Fortin *et al.* [31] in a physical model designed to investigate the behaviour of bubbles underneath the carbon anodes of the Hall process for aluminium production. When the electrode was twisted a few degrees from horizontal, bubbles forming uniformly under the simulated electrodes coalesced to produce a large bubble "front" that moved across the surface and scavenged other bubbles in its path. The process was repeated at a frequency of 1 – 3 Hz when the gas evolution rate was equivalent to 1 A/cm$^2$.

## 4 BUBBLES DETACHMENT

Detachment as the final phase in the physics of gas evolution has also been subject of both theoretical and experimental studies. The bubbles are found to detach once the surface adhesive forces, related to bubble contact angles, can no longer restrain them [32]. In contrast to these equilibrium measurements, studies of the dynamics of gas evolution showed that the bubble formed by the coalescence of two large bubbles would jump off the electrode and sometimes even return [15]. Other authors [27,28,33-35] have also reported that bubble coalescence often precedes their detachment from the electrode surface. It was concluded that the expanding boundaries of the new bubble mechanically forced it off the electrode. It was further speculated that bubbles' movement towards the electrode could be affected by electrostatic forces on a moving bubble or by surface forces varying with concentration.

Related to the detachment of bubbles is their mobility on electrode surfaces. Investigation of the forces holding a drop on an inclined plane led to conclusion that drops holding onto these surfaces are a result of the contact angle hysteresis, the adherence between bubbles' advancing and retreating contact angles [36]. The main cause of hysteresis is roughness of the surface. Later contribution of the same authors [37] valid in the case of gas bubbles as well, established criteria for determining whether or not a bubble should move on an inclined surface. Figure 2, taken from Dussan [37], illustrates the relation between the hysteresis angle, the contact angle, and the volume of the largest bubble that sticks to a surface facing downward and inclined by $\gamma$ degrees to the horizontal and how they determine the minimal and maximal size of electrolytically produced bubbles in aqueous media. With increasing the contact angle and keeping the other parameters constant, increase of bubble size can be observed.

FIG 2

# 5 OPERATING PARAMETERS

Bubble growth and consequently its final size are determined by several parameters including electrode material, current density and different additives. Venczel [38] explored bubble growth on different electrode substrates, namely graphite, iron (Fe), copper (Cu) and Pt as well as on glass plates with thin layers of Pt, chromium (Cr), Ni and gold (Au). Bubble growth on Pt was uniform leading to bubbles coalescence and formation of large bubbles. On the contrary, bubbles were observed to detach from Cu and Fe electrode substrates before reaching a size for coalescence, resulting in small bubbles. Ibl and Venczel [39] confirmed that the size of bubbles depends on experimental conditions and also observed that bubbles evolved on Pt were much larger than those evolved on Cu. Furthermore, it has been established that current density affects bubble size, but there are still some disagreements in the literature about its exact effect. Janssen and Hoogland observed that higher current density values provoked bubbles' coalescence [16]. Their explorations of gas evolution in alkaline media on horizontal and vertical Pt discs [40] showed that all bubbles but $H_2$ increased with increase of current densities (i > 10 mA/cm$^2$); this was attributed to the more frequent coalescence of bubbles. As seen in Figure 3, the size of $H_2$ bubbles produced in alkaline medium is not influenced by current density and they are the smallest ones. On the other hand, $O_2$ bubbles evolved in the same medium rapidly grew for current density values higher than 30 mA/cm$^2$. Some other authors have confirmed the increase of bubbles' size with increasing current density [41,42]. Quite the contrary, Venczel [38] claimed that bubble size decreased with the increase of current density. Up-to-date there are still some disagreements on the exact influence of current density on bubble size.

FIG 3

Besides the electrode material and current density, additives are another parameter believed to influence bubble behaviour and size. Studies showed that presence of additives such as gelatine, glycerine and β-naphthochinolin in the electrolyte solutions results in evolution of smaller bubbles in most cases with formation of frothy mixture being observed [38]. It is believed that additives reduced the ratio of the bubble diameter to the contact diameter by half and increased the electrode wettability. This resulted in the presence of a thick film of electrolyte between the gas and the electrode that is less adhesive than thin films. Still, the forces holding the bubbles to the electrode were most likely weakened proportionally to the decrease of both the perimeter of the contact area and the contact angle. Moreover, it is speculated that the inhibitors stabilise the bubble interfaces and avert their coalescence on the electrode.

In addition to electrode material, current density value and presence of additives in the electrolyte, other process parameters have been reported to influence bubble size and behaviour. It was found that the bubble size decreased with increasing flow rate in the interelectrode area [42] with significantly higher current densities being achieved in comparison to those in stagnant electrolyte. Furthermore, electrode orientation and configuration are also expected to affect bubble size [43].

A more comprehensive study on the effect of the operational parameters such as position, diameter and material of the electrode, nature of gas evolved, temperature, pressure and KOH concentration, was published by Janssen and Barendrecht [44]. They determined the electrolytic resistance of solution layers of KOH at the gas-evolving electrodes, by the alternating current impedance method.

The investigation was restricted to electrodes of small diameter. $\Delta R^*/i$ curves were obtained, where $\Delta R^* = (R - R_0)/R_0$, R being the ohmic resistance of the solution, and $R_0$ the value of R at i = 0. All log $\Delta R^*$ / log i curves for the $H_2$ electrode were linear. Their slope, b, did not depend significantly on KOH concentration. An observed steep slope at 358 K was caused by a high percentage of water vapour in the hydrogen gas-water vapour mixture. Since $\Delta R^*$ increases with increasing gas void fraction in a solution layer at a gas-evolving electrode, it is likely that the observed decrease of slope b with increasing KOH concentration is due to the fact that KOH concentration strongly affects the gas void fraction.

The solution layer at a gas-evolving electrode can be divided into two layers, *viz.* a "fixed layer" (thickness being the diameter of adhered bubbles) adjacent to the electrode surface, followed by a "diffuse layer". To explain the dependence of $\Delta R^*$, it is important to determine the gas void fraction in the "fixed layer". This determination was not possible for the hydrogen-evolving electrode. Generally, for a hydrogen-evolving electrode, $\log \Delta R^* = a_1 + b \log i$ and $\log \Delta R^* = a_2 - b \log p$, where $a_1$ and $a_2$ are constants depending on the position, height and material of the electrode and on temperature. Pressure, p, and current density have similar effects on $\Delta R^*$. So the constant b strongly depends on the KOH concentration. For the oxygen-evolving electrode, the log $\Delta R^*$ / log i curve was not linear but S-shaped.

This result may be caused by coalescence of oxygen bubbles depending on many factors, such as KOH concentration, pressure, temperature and nature of electrode material. In this case, determination of the gas void fraction in the "fixed layer" can help explaining the experimental log $\Delta R^*$/ log i relations.

For an oxygen-evolving electrode, the gas void fraction was determined as a function of the distance from the electrode surface to explain the experimental $\Delta R^*/ \log i$ relations. The reduced resistance increase $\Delta R^*$ is related to a solution layer with a thickness of $0.125 \pi d_e$ at a gas-evolving disc electrode of diameter $d_e$ and a surface area of $0.25\pi d_e^2$. Almost all the bubbles within this layer are adhered to the electrode surface. So, the contribution of "free" bubbles to the gas void fraction can be neglected. $\Delta R^*$ was calculated using the Bruggeman equation [45]

$$p = p_0 (1 - \varepsilon)^{-3/x} \qquad (5)$$

where p and $p_0$ are the effective and real pressures of the solution layer, $\varepsilon$ stands for the dielectric constant of the media and the exponent $-3/x$ is the asphericity parameter of the spherical bubbles. Both $\log \Delta R^* / \log i$ curves calculated and obtained experimentally agreed very well with each other. Consequently, it was shown that the Bruggeman equation is useful to determine the ohmic resistance of a solution layer containing bubbles of different size and at which each bubble adheres to the electrode surface.

# 6 EFFECT OF GAS BUBBLES ON ELECTROLYTE CONDUCTIVITY

The conductivity of a heterogeneous medium, an important topic in diverse technical problems, depends on three characteristics: the ratio of the conductivity of the dispersed and continuous phases, the gross volume fraction occupied by the dispersed phase, and its state of aggregation. Depending on whether its conductivity is greater or lesser than the surrounding medium, the dispersed phase can enhance or retard transport. Volume fraction is often the only parameter other than the aforementioned ratio which appears in equations predicting the effect of a given dispersed phase in the overall conductivity. By state of aggregation we mean both the shape of the dispersed phase, which can range from spheres to cylinders to planes, and its distribution in the system's container.

Meredith and Tobias [46] have written an important review of the role these three characteristics play in the conductivity of heterogeneous media.

The use of any equation describing the conductivity of heterogeneous media generally requires accuracy and convenience. The equation must predict accurate values, be useful over the whole range of void fraction, f, and have a form as simple as possible. There are equations in the literature which satisfy many of these criteria: Maxwell's equation [47], Bruggeman's equation [45], Tobias Distribution Model [48] and Prager's equation [49], shown in equations (6 – 9), respectively.

$$K_m = (1-f)/(1+f/2) \qquad (6)$$

$$K_m = (1-f)^{1.5} \qquad (7)$$

$$K_m = 8(2-f)(1-f)/(4+f)(4-f) \qquad (8)$$

$$K_m = 1 - 3f/2 + 0.5f^2 \qquad (9)$$

$K_m$ is the ratio of the conductance with the dispersed phase present to the conductance in the absence of the dispersed phase. Maxwell's equation (6) [47] is a fundamental result in the theory of heterogeneous conductivity. Prager [49] has improved on Maxwell's result in the concentrated range of void fraction by applying the principle of minimum entropy to obtain bounds on the diffusion coefficient of a solute in a suspension of solid particles. Since the diffusion rate ratio can be considered analogous to the conductivity ratio, one may use his equation (9), exact for a suspension of spheres, to estimate the conductivity of heterogeneous electrolyte. The other two equations were derived for electrolytes containing a distribution of bubble sizes. Meredith and Tobias [48] derived equation (8) for electrolytes containing two sizes of bubbles while Bruggeman [45] treated a pseudo-continuous distribution of bubble sizes by accumulating the contributions of a range of bubble sizes (equation 7). We call it pseudo-continuous because, in principle, each size fraction must be very different from each of the other sizes. In this work we compare the predictions of these four equations to conductivity data spanning the range of void fraction and show that, for many applications, these general equations reliably estimate the heterogeneous conductivity of random dispersions of dielectric spheres. The data are taken from several sources [50-56], the earliest of which appeared in 1926 [50]. Equations (6-9) are compared to the data in Figure 4, although the data follow a similar trend, there are important differences among them. The results of Meredith [55] and DeLaRue [53] follow the same line at low void fractions but, beginning between void fractions of 0.1 and 0.2, DeLaRue's conductivities of dispersions containing spheres of various sizes are consistently lower than Meredith's conductivities and DeLaRue's own conductivities of dispersions of monosized spheres. We conclude that there are three distinct cases in the problem of the conductivity of dispersion of spheres. The first case, dilute dispersions, obeys Maxwell's equation (6) and is independent of the size distribution. The second case, concentrated dispersions of spheres of various sizes, obeys Bruggeman's equation. The third case, concentrated solutions of monosized spheres, is intermediate between the first two in the sense that its conductivities are lower than those predicted by Maxwell's equation and higher than those predicted by Bruggeman's. The distribution model of Meredith/Tobias equation (8) and Prager's equation (9) predict nearly equally accurate values for this case.

FIG 4

These equations and experiments concerned bubbles contained in bulk electrolyte far from electrode surfaces. Sides and Tobias [58,59] theoretically and experimentally investigated the effects of bubble layers on electrodes and concluded that, although the behaviour of the conductivity is definitely different from that in the bulk, the equations derived for the bulk give reasonable estimates if one is careful to apply them to only the region near the electrode where the bubble layer exists. Lanzi and Savinell [60] showed that differences appear at bubble layer void fractions above 0.5.

Apart from these theoretical investigations and experiments on the static models, several authors have studied dynamic gas evolving electrodes. Hine *et al.* [61,62] reported the results of experiments on large-scale vertical electrolysers and concluded that Bruggeman's equation (7) described the results. Sigrist *et al.* [63] also examined electrodes producing bubbles but found that Maxwell's equation best fit their data. Janssen *et al.* [64] investigated ohmic losses in water electrolysis resulting from the presence of bubbles and proposed a correlation to describe the added resistance. Kubasov and Volkov [65], investigating bubble layers formed on graphite electrodes during chlorine evolution, measured voltage losses as high as three volts. Takata *et al.* [66] also worked on chlorine evolving electrodes. In 1956, Hine *et al.* [67] discussed the anode shift potential caused by the local current density exceeding the superficial current density because the bubbles screen the electrode surface. Janssen and Barendrecht [44] investigated the electrolytic resistance of solution layers at hydrogen and oxygen evolving electrodes. Finding that the equations derived for bulk electrolytes worked well, they confirmed the conclusion of Sides and Tobias [59].

Decreasing the local conductivity of the electrolyte, gas evolution can affect macroscopic current distributions. A classic paper in this subject was written by Tobias [68] who analysed the current distribution in vertical electrolysers with evolution of bubbles. Others [61,62,69-71] have performed experiments on the current distribution in such systems.

# 7 MASS TRANSFER AT GAS-EVOLVING ELECTRODES

Electrolytically evolved bubbles enhance heat and mass transfer to gas evolving electrodes because the growing and detaching of gas bubbles mix the electrolyte near the surface with electrolyte in the bulk. For example, Roald and Beck [72] in 1951 noted that the rate of magnesium dissolution in acid electrolyte was controlled by transport of acid to the electrode surface and that stirring by the evolved hydrogen bubbles increased the rate. Evolving gas bubbles can accelerate mass transfer to rates achieved by only intense mechanical stirring or flow and is thus very effective where circumstances allow its use.

Phenomena of gas evolution discussed at the outset are responsible for mixing at the electrode. The growing bubbles produce convection as their boundaries expand by diffusion and coalescence. When a bubble detaches, electrolyte must flow to fill the vacancy. The effective density of a heterogeneous mixture of gas and electrolyte is lower than that of bulk electrolyte; hence there is macro-stirring of the electrolyte by the rising gas bubbles (gas lift). Individually, these are complicated phenomena to describe and it seems impossible to mathematically analyse the flows that are a result of them all; thus, engineers have constructed theories which explain and correlate the experimental results and patterns. Three well-known schools of thought in Switzerland, Netherlands and Germany may be identified from research published over the last fifty years.

The first two schools began with Venczel's dissertation in Zurich in 1961 [73] in which he investigated transport of ferric ion to an electrode evolving hydrogen gas from 1 M $H_2SO_4$. The author observed increased mass transfer with the onset of gas evolution. Ibl and Venczel [39], reporting the results in 1970, attributed the mass transfer enhancement to the reactant's diffusion over short distances from fresh electrolyte brought to the surface after bubble detachment. Surface renewal and penetration theory were used as the basis of a derivation predicting a dependence of the mass transfer rate on the square root of the gas evolution rate. Exponents varying from 0.4 to 0.6 were found experimentally by the authors but a significant number of data were grouped around 0.53. Unfortunately, one must know the bubble size at departure and the fractional area of the electrode shadowed by bubbles on the surface to use this model.

The second school of thought was developed at Eindhoven, in the Netherlands. Janssen and co-workers [16,27,40,74,75] investigated mass transfer enhancement in alkaline electrolyte. $O_2$ bubbles can be quite large (200 μm), whereas $H_2$ bubbles are very small (20 μm) [40]. It was found that the mass transfer enhancement caused by $H_2$ bubbles followed a smaller exponent (0.31) of the gas current density than that reported by Ibl *et al.* [39,51] for low current densities and a large exponent at higher current densities. This change in slope occurred at the same gas evolution rate as the change in bubble size of $O_2$ bubbles from around 60 to 200 µm. Janssen and Hoogland [40] related this change in mass transfer enhancement to increased coalescence and correlated the low exponent to a hydrodynamic model in which convective flows toward a detaching bubble account for the mass transfer.

The third school of thought is represented by Vogt [76] who correlated previously reported results with a model considering the primary enhancement a result of the convective flows originated by expansion of the bubble during growth. He adapted mass transfer theory for laminar flow over planes to his system and integrated the resulting equation over the period of bubble growth. His equation appeared to successfully correlate data from other investigators.

Investigators of mass transfer enhancement at gas-evolving electrodes have generally reported Nernst mass transfer boundary layer thickness, $\delta_N$, as a function of gas evolution rate ($cm^3/cm^2s$) or current density. The Nernst boundary layer thickness is given by

$$\delta_N = nFDC/i \tag{10}$$

where n stands for the equivalents per mole, D is the diffusivity and F is Faraday's constant. $\delta_N$ is essentially the reciprocal of a mass transfer coefficient divided by the molecular diffusivity. Figure 5 is a plot of the results of Ibl *et al.* [77] on mass transfer enhancement in acid solutions. It may be observed that the boundary layer thickness decreases with the increase of the gas evolution rate. At the highest gas evolution rates, corresponding to 10 A/cm$^2$, the boundary layer thickness is of the order of 1 µm which is quite thin. The Nernst boundary layer thickness is a simple characteristic of the mass transfer rate and is usually expressed by

$$\delta_N = ai^b \tag{11}$$

taking into account fluid viscosity, kinematic viscosity and the diffusion coefficient, being the exponent b the centre of differences between the schools of thought, taking on values from -0.25 to -0.87 [51].

FIG 5

Figure 6 was taken from Janssen and Hoogland [40] and it is the basis of their hydrodynamic theory of mass transfer in the absence of coalescence. The slope change, around 30 mA/cm$^2$, corresponds to a change in bubble size for the $O_2$ bubbles evolved in basic solution. This was considered an indication of the onset of coalescence between bubbles on the electrode. Vogt [76] correlates results from many authors with the equation

$$Sh = 0.925 Re^{0.5} Sc^{0.487} \qquad (12)$$

where Sh is the Sherwood number defined as kd/D with k being the mass transfer coefficient, Sc is the Schmidt number, Re is the Reynolds number being defined by $V_G d/v$, $V_G$ is the gas evolution rate, d is the bubble breakoff diameter, and v is the kinematic viscosity.

FIG 6

Regardless of the differences between the three main schools of thought on mass transfer at gas evolving electrodes, some issues have reached consensus among the several investigators. Among them is the fact that gas evolution is considered an effective means of enhancing transport. It is also generally accepted that roughness of the electrodes on the order of the boundary layer thickness does not strongly influence the amount of acceleration. A slight dependence on electrode orientation and on the length of a vertically oriented electrode is also recognized, but as secondary effects [78-82].

## 8 RECENT DEVELOPMENTS AND FUTURE TRENDS

Electrolytic gas evolution has been an interesting field for researchers and great progress has been achieved, but there are still challenging questions to be answered [83-93]. The majority of the studies have focused on miniature or micrometer-sized bubbles that are formed at the electrodes' surface and subsequently detach from it. Recently, electrochemically generated nanobubbles, which work as embryos for the bubbles, have been receiving a great deal of attention [94-97]. These nanobubbles behave differently from macroscopic ones, having contact angle lower than expected from Young's law, being stable against violent decompression, and for much longer than theoretically predicted [97]. They were found to grow into microbubbles before detaching from the substrate [95].

However, the nucleation of gas bubbles at the electrode surface still needs clarification. The phenomena of growth, coalescence, and detachment of bubbles require further enlightenment. Although the conductivity of electrolytes containing gas bubbles is theoretically understood, these effects in large-scale systems, such as in industrial electrolysers, are yet to be fully clarified [98]. The voltage balance at the gas evolving electrode is also not clear. The enhancement of mass and heat transfer at gas evolving electrodes have been widely investigated, but more comprehensive semi-empirical analysis is still required.

Furthermore, the discovery of surface nanobubbles opened a new field of research and numerous questions have been raised, including issues regarding their nature and unexpected long stability, which are important aspects for electrolytic gas evolution.

## 9 CONCLUSION

Gas bubbles formed at the solid/liquid interface are a classical phenomenon that is of great importance in various processes, including electrolytic gas evolution. Many studies have been devoted to the topic but there are still important questions and areas to be explored.

The details of bubble formation and the effects of the bubbles presented in this brief analysis based on the pioneering works of Ibl, Janssen, Sides, Tobias, Venczel, Vogt, and many others, are nothing but the microscopic aspects of a phenomenon that affects the macroscopic behaviour of electrochemical cells, namely those developed for the electrolytic hydrogen production. Going deeper into the ultramicro- and nanoscopic aspects of bubble formation at the tertiary system - water, highly concentrated gas and solid electrode - one comes across with surface nanobubbles, whose existence at the solid/liquid interface has a relevant impact on the nucleation and growth of gas bubbles.

# Acknowledgements

FCT, the Portuguese Foundation for Science and Technology, is gratefully acknowledged for funding project *Functional materials for electrolytic hydrogen production* (PTDC/SEN-ENR/121265/2010) and L. Amaral thanks for a research grant within this project. D.M.F. Santos and B. Šljukić would also like to thank FCT for postdoctoral research grant (SFRH/BPD/63226/2009) and SFRH/BPD/77768/2011, respectively.

# Figure captions

FIG. 1: Documentation of specific radial coalescence, from Sides and Tobias [30]. The indicated bubble at time zero receives the four bubbles around it sequentially and visibly grows. One may observe this same effect around other bubbles in the sequence. Conditions: 10,000 frames per second, oxygen evolution, 298 K, no forced convection, 3 wt. % KOH, 500 mA/cm$^2$.

FIG. 2: Dimensionless volume of largest bubble or drop that adheres to an inclined surface; graph from Dussan [37].

FIG. 3: Bubble breakoff diameter as a function of current density for a Pt gas-evolving electrode (horizontal) in various solutions [40].

FIG. 4: Comparison of equations (6-9) predicting the reduced conductivity dispersions of spheres of unequal sizes with data over the whole range of void fraction [57].

FIG. 5: Thickness of diffusion layer at $H_2$-evolving horizontal Pt-electrodes of different roughness [77].

FIG. 6: Nernst boundary layer thickness for horizontal gas-evolving Pt electrode as a function of current density in various solutions [40].

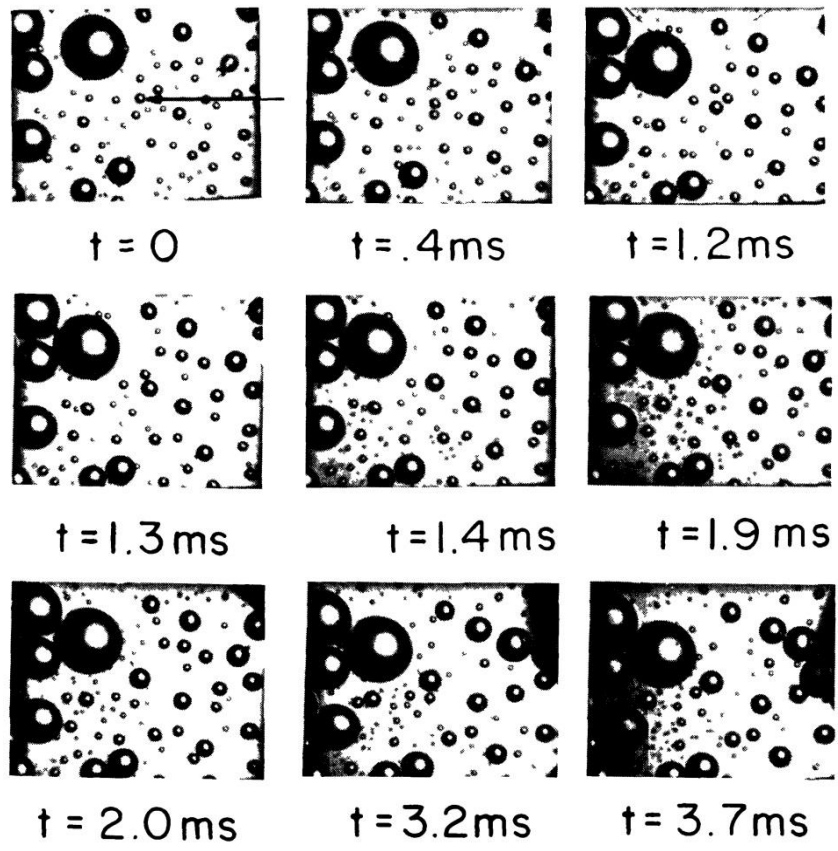

Fig. 1

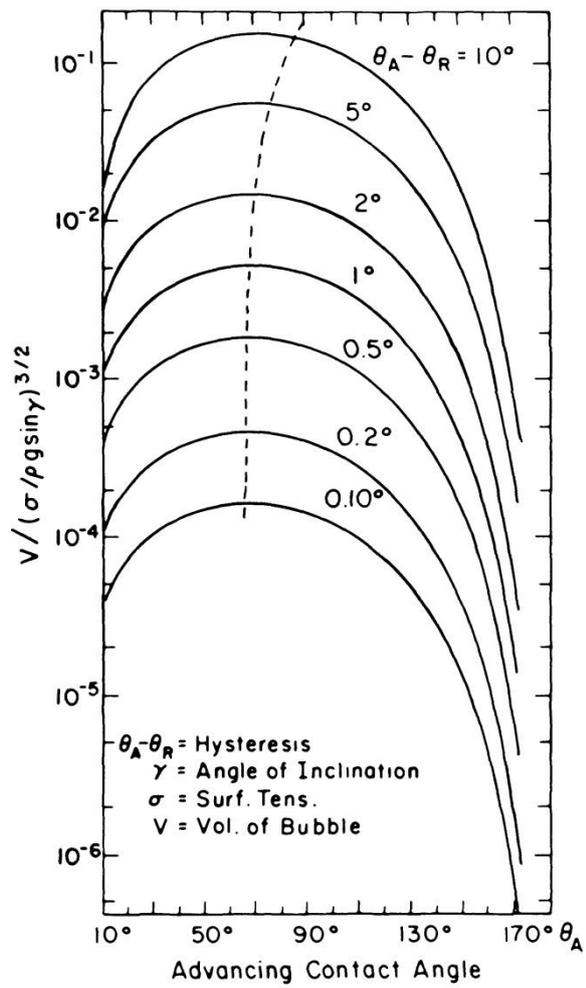

Fig. 2

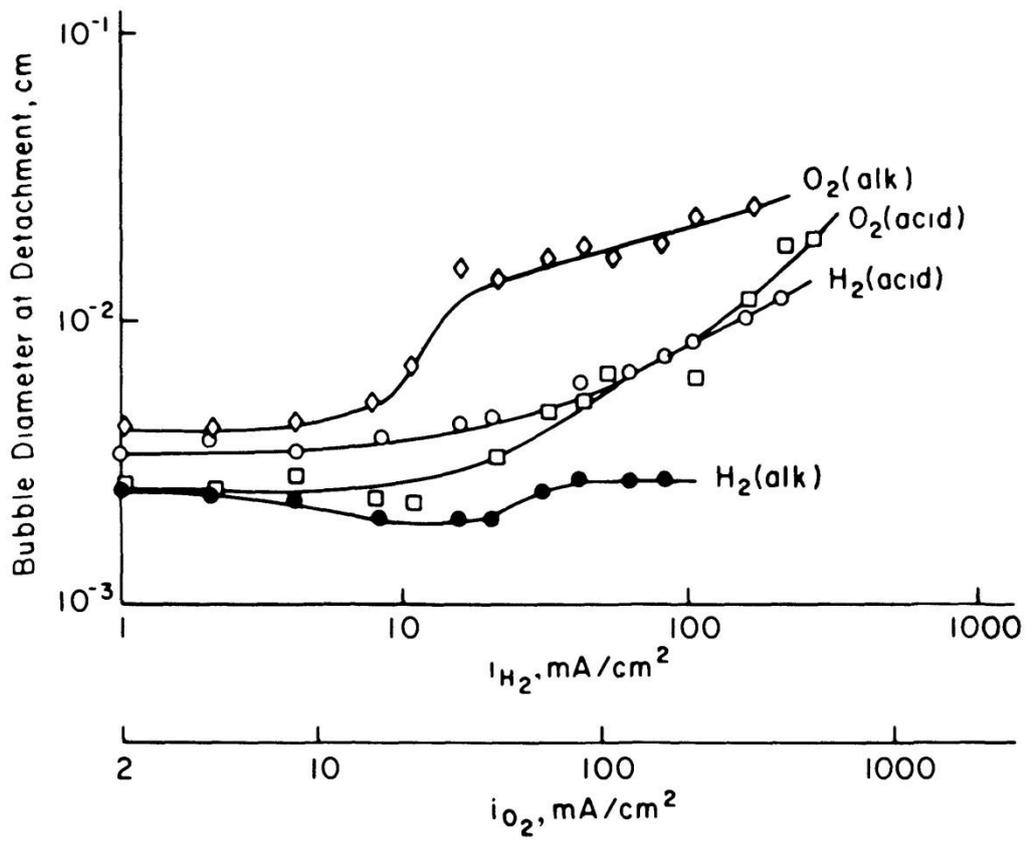

Fig. 3

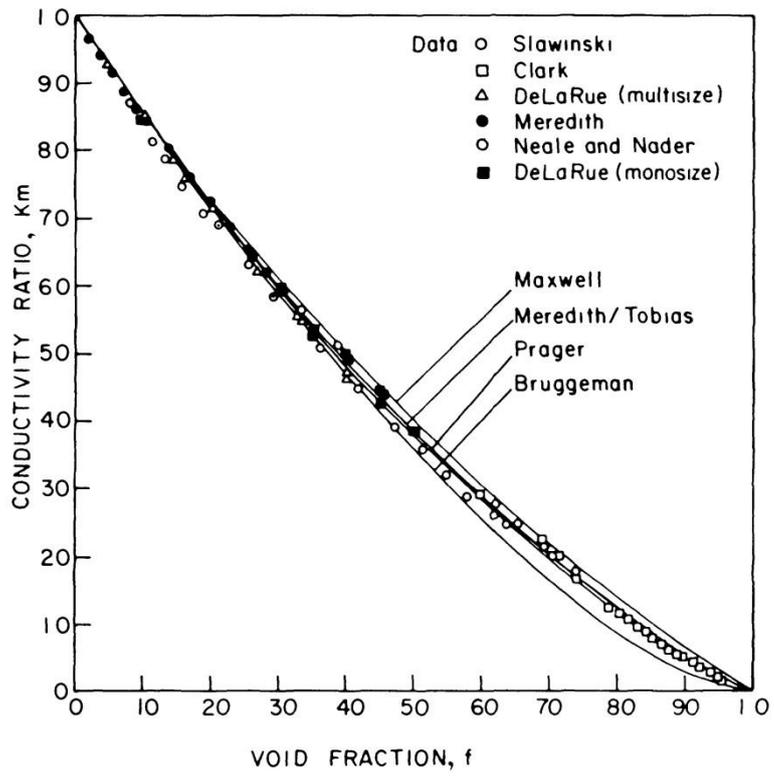

Fig. 4

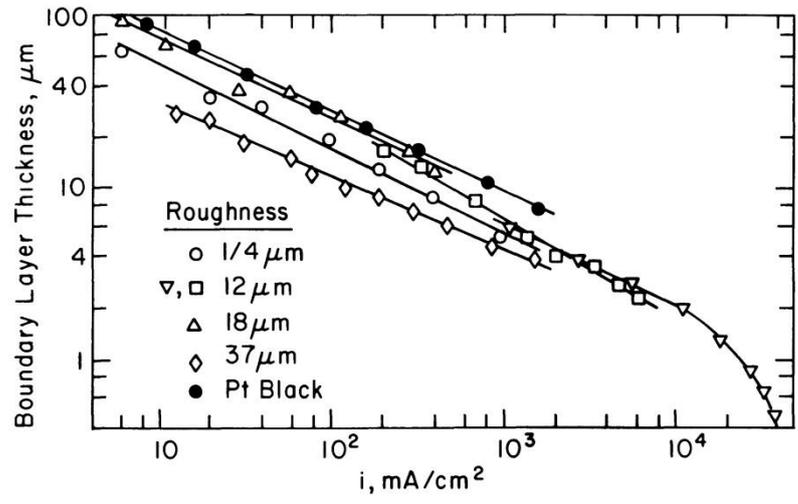

Fig. 5

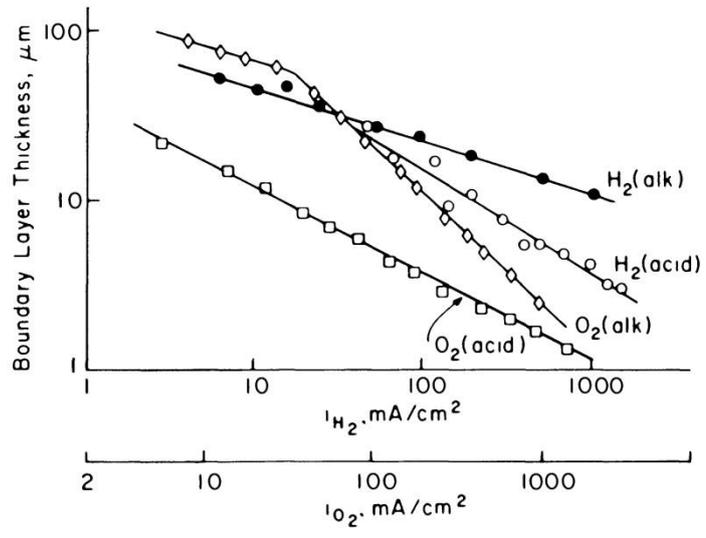

Fig 6.